\documentclass[preprint,12pt]{elsarticle}

\usepackage{amssymb, amsmath}

\usepackage{subfigure}
\usepackage{gensymb}

\journal{}

\begin{document}

\begin{frontmatter}



  \title{Radial hopper flow prediction using a constitutive model with
    microstructure evolution}


\author[label1]{Jin Sun}
\author[label2]{Sankaran Sundaresan}

\address[label1]{ School of Engineering, University of Edinburgh,
  Edinburgh EH9 3JL, UK } 

\address[label2]{ Department of Chemical and
  Biological Engineering, Princeton University, \\Princeton, NJ 08544,
  USA}

\begin{abstract}
  We present theoretical predictions of granular flow in a conical
  hopper based on a continuum theory employing a recently-developed
  constitutive model with microstructure evolution by Sun and
  Sundaresan~\cite{Sun:11a}.  The model is developed for strain
  rate-independent granular flows. The closures for the pressure and
  the macroscopic friction coefficient are linked to microstructure
  through evolution equations for the coordination number and
  fabric. The material constants in the model are functions of
  particle-level properties. A salient prediction is the variable
  stress ratio along the flow direction, in contrast to the constant
  ratio employed in some widely-used plasticity theories, but
  supported by results obtained from discrete element simulations. The
  model permits direct interrogation of the influence of
  particle-particle friction as well as normal-stress differences on
  the stress distribution and discharge rate. Increasing particle
  friction leads to higher stress ratios, but lower normal stress and
  flow rates, while considering normal-stress differences results in
  the opposite effects.
\end{abstract}

\begin{keyword}

  Granular flow \sep constitutive model \sep microstructure
  evolution\sep conical hopper
\end{keyword}

\end{frontmatter}


\section{Introduction}
\label{}
The flow of granular materials in bins and hoppers is of great
practical interest in connection with the handling and transportation
of bulk solids such as sand, coal, ore and grain.  It has been
extensively studied for several decades and generated a significant
number of publications with earlier results summarized in
monographs~\cite{Drescher:91a, Nedderman:92a}, and review
articles~\cite{Nedderman:82a,Tuzun:82a}. With advances in experimental
techniques and computational power, recent results using particle
image velocimetry (PIV)~\cite{Langston:97a, Bohrnsen:04a, Choi:05a,
  Garcimartin:11a, Sielamowicz:11a} and discrete element method
(DEM)~\cite{Langston:95b, Potapov:96a, Zhu:06c, Ketterhagen:09a,
  Rycroft:09a, Hilton:10a, Balevicius:11a} have shed light on flow
kinematics and on particle-property effects on the flow behavior. The
understanding also drives the development of continuum
theories~\cite{Kamrin:07a, Wojcik:09a, Kamrin:10a}, which are
necessary for predicting industry-scale hopper flow.

In spite of the extensive research, a continuum model capable of
accurately predicting realistic hopper-flow behavior is still under
development. For example, the challenges faced in the constitutive
modelling of hopper flow were described in~\cite{Drescher:98a};
deficiencies of a Mohr-Coulomb plasticity theory were summarized
in~\cite{Kamrin:07a}. The deficiencies of current models are also
reflected in the fact that widely different predictions of
well-defined silo flow were made using different
models~\cite{Rotter:98a}. Precisely due to these difficulties, the
hopper-flow problem also serves as a stringent benchmark case to test
newly-developed theories for granular materials, such as the
stochastic flow-rule~\cite{Kamrin:07a}, non-local
hypoplasticity~\cite{Wojcik:09a} and elasto-plastic
models~\cite{Kamrin:10a}.

In this paper, we employ a new constitutive model with microstructure
evolution~\cite{Sun:11a} to predict flow in a conical hopper. We
utilize the radial solution, first presented by
Jenike~\cite{Jenike:61a}, to reduce the number of spatial variables to
one by the inherent symmetry. Such a solution, albeit being rather
simplified, helps to elucidate the essential capabilities of the
constitutive model and prevents obscuring the features by complexities
in boundary conditions and numerical methods. We will demonstrate that
the model is capable of predicting a varying stress ratio along the
flow path, which is realistic but elusive in previous theories. We
will show how the particle properties and microstructure are linked to
the continuum variables, which enables studying the role of such
microscopic parameters at a much lower computational cost. We will
also compare the predictions to DEM simulation results and the
so-called hour-glass theory (HGT)~\cite{Savage:81a, Jackson:83a,
  Nedderman:92a} that is also a radial-flow solution, but employing
the Mohr-Coulomb plasticity theory.

The remaining part is organized as follows; the hopper-flow model is
described in section~\ref{sec:model} detailing the governing
equations~(\ref{sec:gov}) and the constitutive
model~(\ref{sec:const}). The results from solving the model are
presented and analyzed in section~\ref{sec:res}, leading to the
conclusions and future work in section~\ref{sec:conc}. Regarding
notation, we employ lightface italics for scalars, the boldface
regular font for vectors and the San Serif font for the second-order
tensors.

\section{Hopper flow modeling \label{sec:model}}

\subsection{Flow conditions and governing equations \label{sec:gov}}
\begin{figure}
  \centering
     \subfigure[]{
  \includegraphics[width=1.5 in]{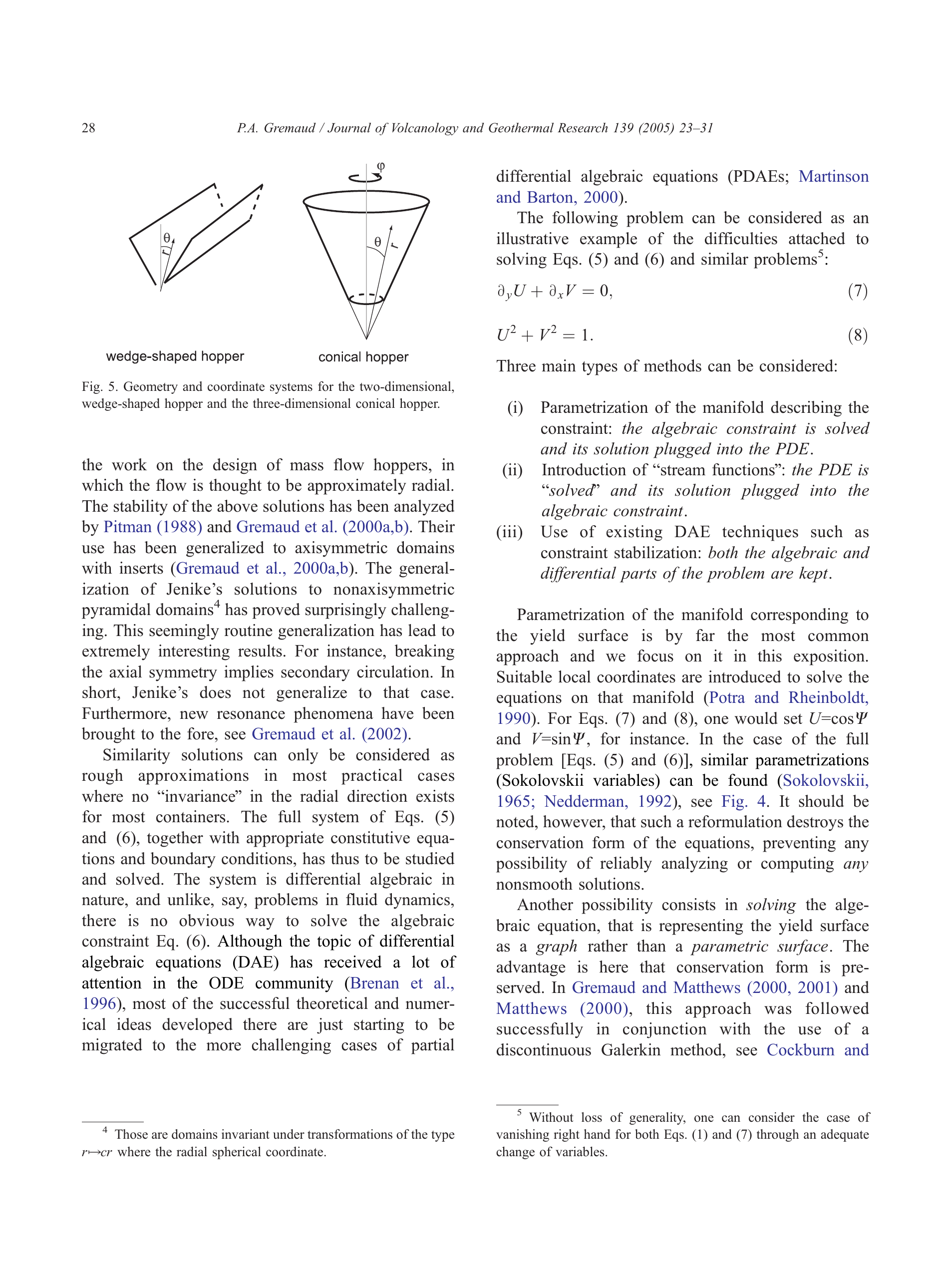}}
      \subfigure[]{
  \includegraphics[width=1.5 in]{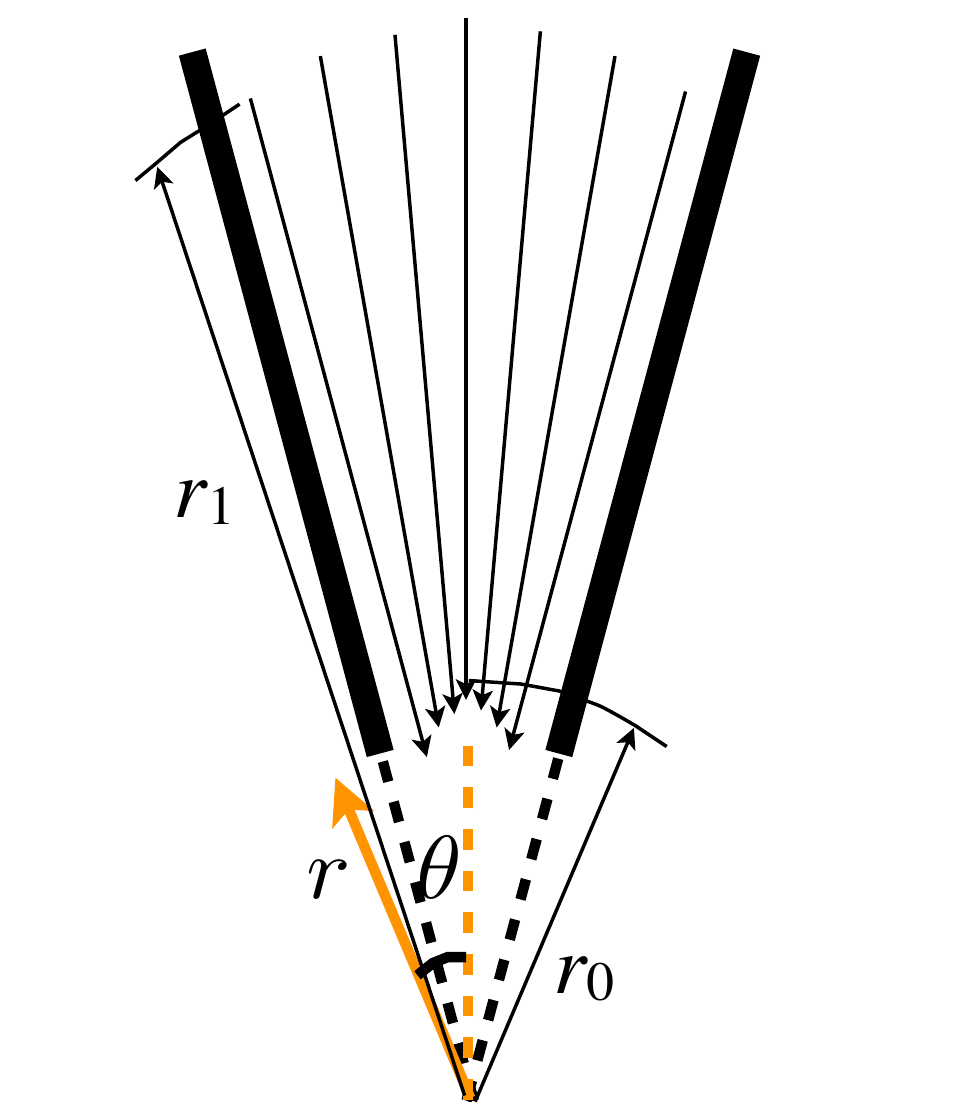}}
\caption{The schematic diagram of the geometry and the spherical
  coordinate systems for a conical hopper showing (a) a
  three-dimensional view and (b) a $r$-$\theta$ cross-section view
  with arrows pointing to the virtual apex indicating the radial
  flow. \label{fig:Geometry}}
\end{figure}

We consider an incompressible cohesionless granular material of mass
density $\rho$ and volume fraction $\phi$ discharging from a
narrow-angled conical hopper of half angle $\alpha$, whose geometry is
sketched in Fig.~\ref{fig:Geometry}. We use a spherical coordinate
system (($r$, $\theta$, $\varphi$) see Fig.~\ref{fig:Geometry}(a))
with origin at the virtual apex of the hopper. The walls are assumed
to be rigid and smooth. Since the angle $\alpha$ is small, gravity is
assumed to be in the radial direction. By axial symmetry, the shear
stress $\sigma_{r\theta}$ is reasonably assumed to be zero. We further
assume that the velocity is radial and independent of $\theta$ and
$\varphi$ coordinates.

With these conditions, the continuity equation for this incompressible
granular flow is reduced to
  \begin{equation}
    \label{eq:continue}
    v_r = - \frac{V}{r^2} \text{,}
  \end{equation}
  where $V$ is a positive flow rate independent of $r$. This flow rate
  is related to the discharge mass flow rate $W$, as $W =
  2\pi\rho\phi(1-\cos\alpha)V$. We then obtain a dimensionless
  momentum conservation equation utilizing Eq.~\ref{eq:continue}
\begin{equation}
  \label{eq:mom1}
  \frac{d\hat{\sigma}_{rr}}{d\hat{r}} =
  \frac{2\phi\hat{V}^2}{\hat{r}^5} +
  \frac{\hat{\sigma}_{\theta \theta} + \hat{\sigma}_{\varphi \varphi} -2\hat{\sigma}_{rr}}{\hat{r}} - \phi \text{,}
\end{equation}
where $r$ is scaled by the radius of a free-fall arch $r_0$ (see its
location in Fig.~\ref{fig:Geometry}(b)), i.e. $\hat{r} = r/r_0$; and
the normal stresses $\hat{\sigma}=\sigma/(\rho gr_0)$ and $\hat{V} =
V/\sqrt{r_0^5g}$. Solution of Eq.~\ref{eq:mom1} with appropriate
boundary conditions predicts the radial distributions of velocity and
stress as well as the flow rate. However, this equation is not closed
and a constitutive model for the stress is necessary. We will present
the model employed in this study in the next section.

\subsection{Constitutive model}
\label{sec:const}
We briefly present the constitutive model in its general form for
completeness. Readers are referred to~\cite{Sun:11a} for details on
the derivation and characteristics of the model. In this model, the
stress is expressed as
\begin{equation}
\label{eq:augment_stress_a6}
\mathsf{\boldsymbol\sigma} = p\mathsf{I} -
p\eta\hat{\mathsf{S}}+
a_3p(\mathsf{A}\hat{\mathsf{S}} + \hat{\mathsf{S}}\mathsf{A}
-\frac{2}{3}(\mathsf{A:\hat{S})I})+a_4p(\mathsf{A}
-\frac{1}{2}(\mathsf{A}:\hat{\mathsf{S}})\hat{\mathsf{S}})\text{,}
\end{equation}
where $p$ is the pressure or mean normal stress, $\hat{\mathsf{S}}
:=\mathsf{S}/\sqrt{\frac{1}{2}\mathsf{D}^{\mathrm{T}}:\mathsf{D}}$
with $\mathsf{S}$ and $\mathsf{D}$ as the deviatoric and full strain
rate tensors, respectively, i.e.~$\mathsf{S}
=\mathsf{D}-\frac{1}{3}\mathrm{tr}(\mathsf{D})\mathsf{I}$, $\eta$ is a
scalar macroscopic friction coefficient and $\mathsf{I}$ is the
identity tensor. The last two terms in Eq.~\ref{eq:augment_stress_a6}
model the normal stress differences (NSD) in simple shear flow,
without which the model predicts equal normal stresses for such
flow. The pressure and the macroscopic friction coefficient are linked
to coordination number $Z_2$ and fabric tensor $\mathsf{A}$ using
\begin{equation}
  \label{eq:press}
     pd/k = (a_1+ a_2|\mathsf{A}|)(Z_2-Z_c)^{\alpha} \text{,}
\end{equation}
and
\begin{equation}
  \eta = b_1 + b_2\mathsf{A}:\hat{\mathsf{S}}\text{,}
  \label{eq:eta_linear}
\end{equation}
respectively, where $d$ is the particle diameter, $k$ is the particle
normal stiffness, and $Z_c$ a critical coordination number
characterizing jamming. In
Eqs.~\ref{eq:augment_stress_a6}--\ref{eq:eta_linear}, the
coefficients, $a_i$, $b_j$ ($i=1,2,3,4$ and $j=1,2$) and $\alpha$, are
material constants with $a_{1,2}$ and $\alpha$ independent of particle
friction coefficient $\mu$ and the rest as a function of $\mu$. Their
numerical values are given in Table~\ref{tab:numbers}.

The fabric evolution is modeled as
\begin{equation}
  \mathring{\mathsf A} = c_1{\mathsf S} + c_2 |\mathsf{D}|{\mathsf{A}} +
  c_3({\mathsf{A:S}}) {\mathsf A} \text{,}
    \label{eq:A}
\end{equation}
where $\mathring{\mathsf{A}} = \dot{\mathsf{A}} + \mathsf{A}\cdot
\mathsf{W} - \mathsf{W}\cdot \mathsf{A}$, with $\mathsf{W}$ as the
spin tensor, $\mathsf{W} = \frac{1}{2} (\nabla \mathbf{v} - (\nabla
\mathbf{v})^{\mathrm{T}})$, and $\dot{\mathsf{A}}$ denotes its
material time derivative. The coordination-number evolution is given
as
\begin{equation}
  \label{eq:Z}
  \dot{Z_2} =d_1(\mathsf{A:S} - 
  \chi |\mathsf{S}|) + d_2 |\mathsf{D}|(f(\phi) -Z_2) +
  d_3 \mathrm{tr}(\mathsf{D}) \text{,} 
\end{equation}
where $\chi = -(c_2 + \sqrt{c_2^2-8c_1c_3})/2c_3$ equals
$\mathsf{A}:\hat{\mathsf{S}}$ for steady simple shear according to
Eq.~\ref{eq:A}, $  f(\phi) = Z_c + \beta_1(\phi-\phi_c)^{\beta_2}$ and
$\mathrm{tr}(\mathsf{D})$ is the trace of $\mathsf{D}$. In Eqs.~\ref{eq:A}  and~\ref{eq:Z}, the coefficients,
$c_i$, $d_i$ ($i=1,2,3$) and $\beta_j$ ($j=1,2$), are material constants modeled independent of $\mu$,
whose values are listed in Table~\ref{tab:numbers2}.

\begin{table}
  \centering
  \begin{tabular}{cccccccccc}
\hline \hline
           &$\alpha$& $a_1$     &   $a_2$         & $a_3$ &$a_4$&
           $b_1$ & $b_2$ & $Z_c$ & $\phi_c$ \\ 
\hline
$x_1$  & 2           &0.0073   & -0.1   & 1.57   &-6    & -0.16  & 1.6     & 1.85 & 0.058 \\ 
$x_2$  &              &&           &  -4.5  &-2    & -6       &-6& -5 & -5 \\ 
$x_3$  &              &&           & -1.7   & 6     &0.16     & -2.9&4.15 & 0.582\\
\hline
    \end{tabular}
    \caption{Numerical values of $\alpha$, $a_1$, $a_2$ and $x_1$-$x_3$ in the fitting
      expression $x_1e^{x_2\mu}+ x_3$, for the material parameters in the
      pressure and $\eta$ equations, and the $Z_2$ evolution
      equation~\cite{Sun:11a}. \label{tab:numbers}}
\end{table}

\begin{table}
  \centering
  \begin{tabular}{c c c c c c c c  }
\hline \hline
    $ c_1$ &$ c_2$ & $ c_3$ &$d_1$ & $d_2$ &$d_3$
    &$\beta_1$&$\beta_2$\\
\hline
    -0.52&-2.8&100 & -45  &5.6 &-40 & 7.5 & 0.5\\
\hline
    \end{tabular}
    \caption{Numerical values of the material parameters in the
      microstructure evolution equations~\cite{Sun:11a}. \label{tab:numbers2}}
\end{table}

Using the assumptions for radial hopper flow in section~\ref{sec:gov},
the strain-rate tensor is diagonal, with $[D_{rr}, D_{\theta\theta},
D_{\varphi\varphi}]=[\frac{dv_r}{dr}, \frac{v_r}{r}, \frac{v_r}{r}]$
giving $\hat{\mathsf{S}} =[\frac{2}{\sqrt{3}},-\frac{1}{\sqrt{3}},
-\frac{1}{\sqrt{3}} ]$. This together with the fabric Eq.~\ref{eq:A}
determines that the fabric tensor is also diagonal,
i.e.~$\mathsf{A}=[A_{rr}, A_{\theta\theta}, A_{\varphi\varphi}]$, with
$ A_{\theta\theta} = A_{\varphi\varphi}$. This means $
-\frac{1}{2}A_{rr} = A_{\theta\theta} = A_{\varphi\varphi}$ since the
fabric tensor is defined as being traceless.  The fabric evolution
equation is thus simplified to
\begin{equation}
  \label{eq:Arr}
   \frac{dA_{rr}}{d\hat{r}} = -\frac{1}{\hat{r}}(2c_1 +
   \sqrt{3}c_2A_{rr} + 3c_3A_{rr}^2) \text{.}
\end{equation}
The boundary condition of $A_{rr} = 0$ at the top surface of radius
$r_1$ (see Fig.~\ref{fig:Geometry} (b) for its location) is applied
for this equation, which is a reasonable assumption of the isotropic
state due to random deposition and lack of substantial shear
deformation at the top.

Substituting the fabric and strain-rate tensors in
Eq.~\ref{eq:eta_linear}, we obtain
\begin{equation}
  \label{eq:eta}
  \eta = b_1 + \sqrt{3}b_2A_{rr}\text{.}
\end{equation}
Similarly with the stress Eq.~\ref{eq:augment_stress_a6}, we have the
stress components
\begin{align}
  \sigma_{rr} &=p\left (1-\frac{2}{3}\mu_b\right )  \text{,} \\
  \nonumber\sigma_{\theta\theta} &=\sigma_{\varphi\varphi}
  =p\left (1+\frac{1}{3}\mu_b\right ) \text{,} 
\end{align}
where $\mu_b$ is a bulk friction coefficient defined as the ratio of
(Von Mises) shear stress to pressure and
\begin{equation}
  \label{eq:mub}
  \mu_b=\sqrt{3}(\eta -a_3A_{rr})\text{.}
\end{equation}
The bulk friction coefficient can thus be calculated from fabric and
particle friction and need not be taken as a constant as in many
plasticity theories. Defining the relation between the stress
components as $K \sigma_{rr} = \sigma_{\theta\theta} =
\sigma_{\varphi\varphi}$, we then get the stress ratio
\begin{equation}
  \label{eq:K}
  K = \frac{1 + \frac{1}{3}\mu_b}{1 -\frac{2}{3}\mu_b}\text{.}
\end{equation}
It can be seen that the relation between the stress components is
derived from the stress equation given the fabric and flow
conditions. We do not invoke the Haar-von Karman
hypothesis~\cite{Nedderman:92a} to reach this result.

With this relation for stress components, the momentum conservation
given in Eq.~\ref{eq:mom1} can be recast as
\begin{equation}
  \label{eq:mom}
  \frac{d\hat{\sigma}_{rr}}{d\hat{r}} =
  \frac{2\phi\hat{V}^2}{\hat{r}^5} +
  \frac{2(K-1)\hat{\sigma}_{rr}}{\hat{r}} - \phi \text{,}
\end{equation}
For this equation, we apply two boundary conditions, assuming a
free-fall arch of radius $r_0$ and stress-free top surface of radius
$r_1$. We note that the free-fall arch concept is echoed by the
force-chain idea in recent granular material research. Such arch can
be seen from the visualization of force chains in the DEM simulation
of a flat-bottom silo~\cite{Zhu:06c}. Equation~\ref{eq:mom} is closed by
the constitutive relations Eqs.~\ref{eq:Arr}--\ref{eq:K} and specific
solutions can be obtained applying the boundary conditions. The
pressure and $Z_2$ evolution equations are redundant for
incompressible flow, as evidenced from the above derivation process.

\section{Hopper flow predictions \label{sec:res}}
\begin{figure}
  \centering
      \subfigure[]{
  \includegraphics[width=2.5 in]{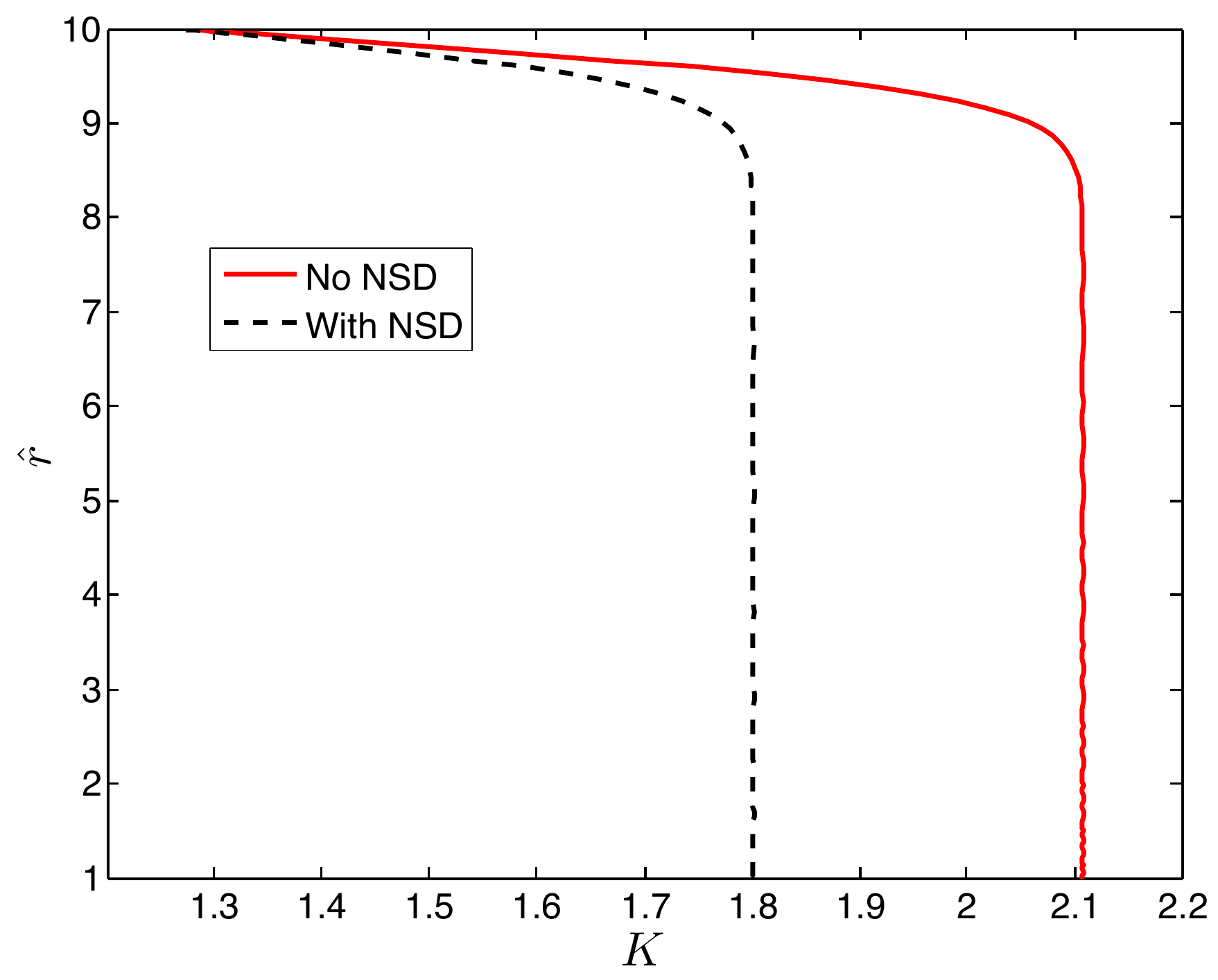}}
      \subfigure[]{
  \includegraphics[width=2.5 in]{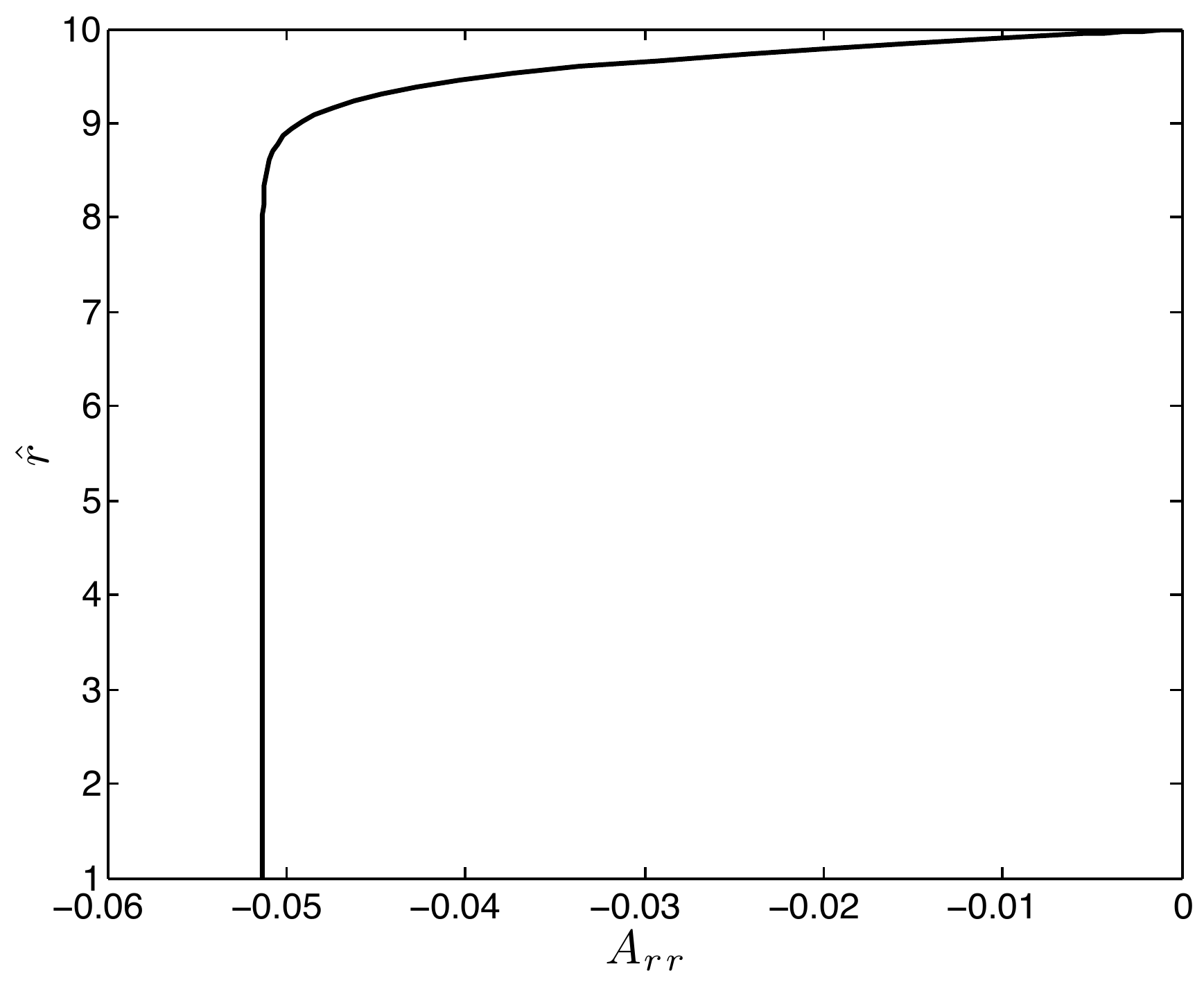}}
      \subfigure[]{
  \includegraphics[width=2.5 in]{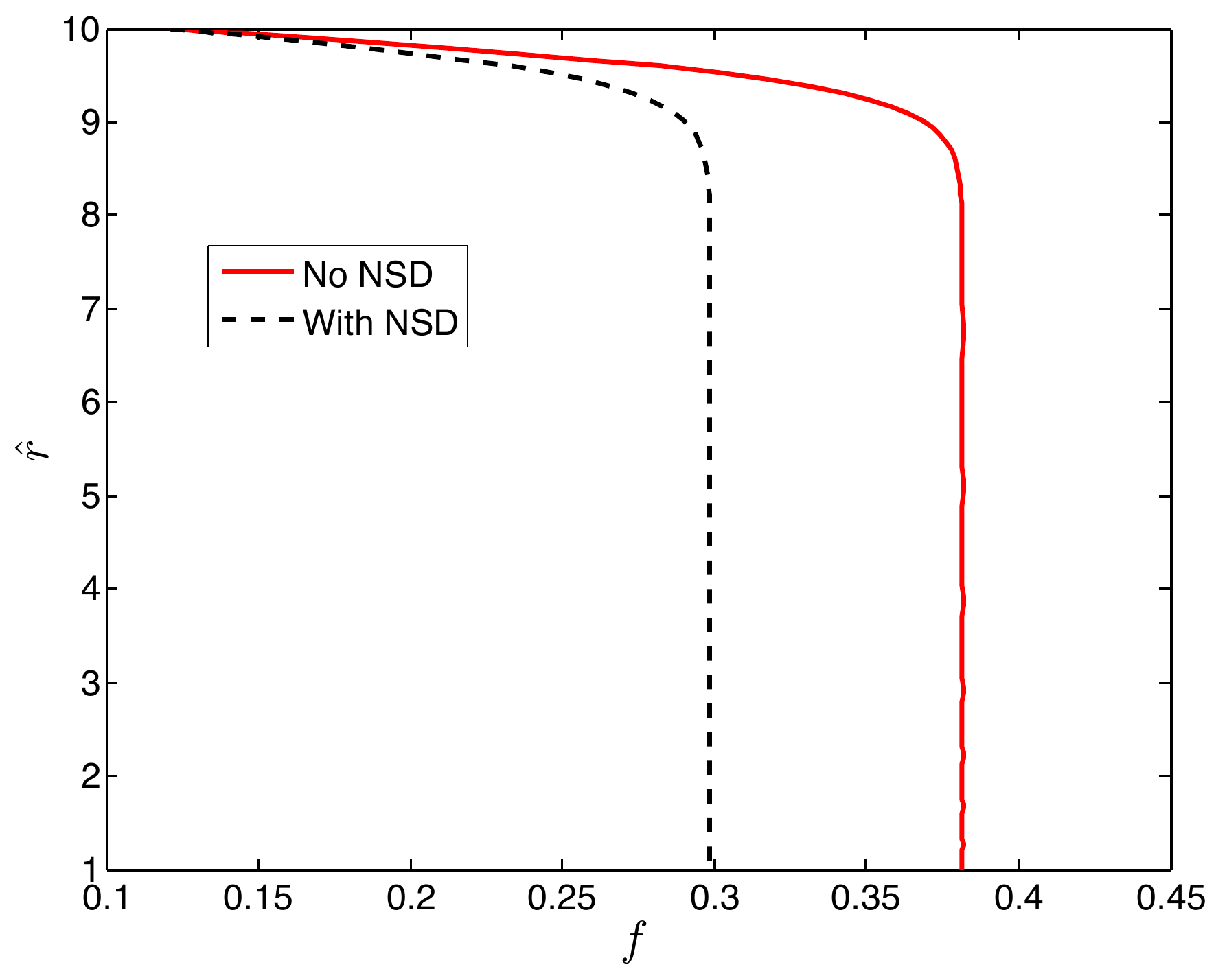}}
     \subfigure[]{
  \includegraphics[width=2.5 in]{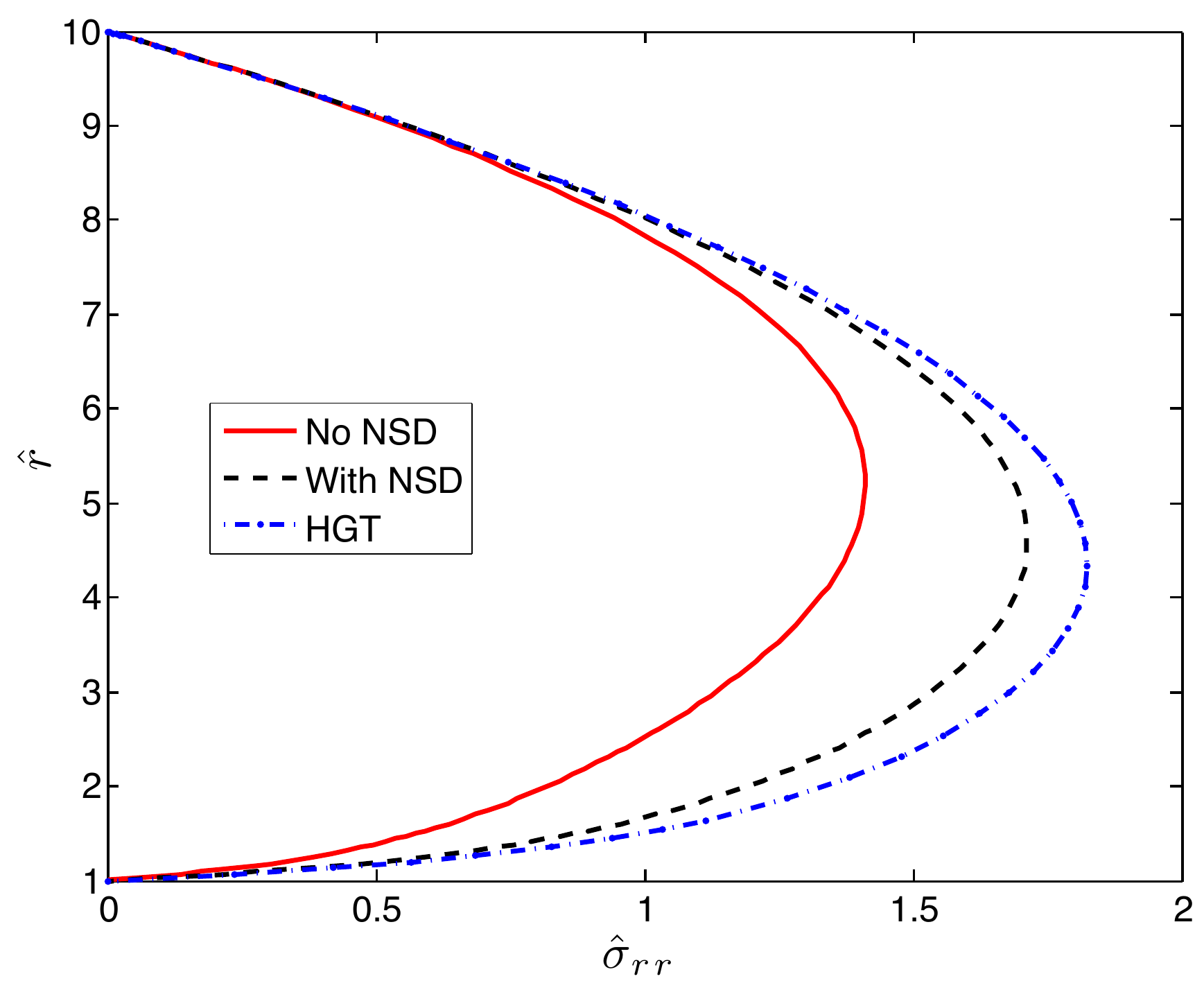}}

\caption{Radial variation of (a)stress ratio, (b) component of the
  fabric tensor, (c) macroscopic friction coefficient and (d) normal
  stress, plotted against the dimensionless distance in the vertical
  axis. The material has particle friction
  $\mu=0.3$. \label{fig:stress_mu03}}
\end{figure}

Solving the radial model obtained in section~\ref{sec:const} for
hopper flow of particles with friction $\mu=0.3$, the stress and
fabric radial distributions are obtained and shown in
Fig.~\ref{fig:stress_mu03}. The present theory predicts that the
normal-stress ratio $K$ increases along the flow direction (see
Fig.~\ref{fig:stress_mu03} (a)), which naturally arises from the
variation of the fabric (i.e.~$A_{rr}$, shown in
Fig.~\ref{fig:stress_mu03} (b)). The value $A_{rr}$ increases along
the flow, indicating microstructure anisotropy increased due to large
deformation. As prescribed in Eq.~\ref{eq:eta}, the increase of
$A_{rr}$ results in the increase in $\eta$ along the flow direction,
which in turn increases the $K$ values according to Eq.~\ref{eq:K}.
The NSD effect further couples the $A_{rr}$ variation to $K$,
reflected in Eq.~\ref{eq:mub}. As seen in Fig.~\ref{fig:stress_mu03}
(a), the formulation with the NSD leads to smaller $K$ values, $1.8$
at the bottom, comparing to $K= 2.1 $ calculated without the NSD. In
contrast, a constant $K$ is prescribed by the Mohr-Coulomb yield
criterion together with the incipient-yield-everywhere assumption,
which was employed in many previous hopper-flow analyses~\cite[more
references therein]{Savage:81a, Nedderman:92a}, and
simulations~\cite{Gremaud:01a}. The essential difference in the
present constitutive model is that the $K$ value, a bulk level
property, need not be specified \textit{a priori}; but is calculated
from the particle-level property, $\mu$, and the microstructure
evolution. Another distinction has to be made from the constant $K$
used in pressure analysis for a static granular column, which appears
to be a reasonable assumption in the static condition, as seen in
Janssen's analysis~\cite{Nedderman:92a} and verified more recently by
an elasticity theory~\cite{Brauer:06a}. This comparison highlights the
fact that the stress state in the flow is different from that in the
static condition.

To compare these continuum results with DEM ones, we calculate an
internal friction coefficient $f$, using the same definition as in a
DEM study of two-dimensional hopper flow~\cite{Potapov:96a}. This
coefficient is related to the stress ratio $K$ as $f
=\frac{K-1}{2\sqrt{K}}$. The value of $f$ increases from $0.121$ at
the top to $0.382$ ($0.300$) at the bottom for the case without (with)
the NSD as plotted in Fig~\ref{fig:stress_mu03}(c). This variation
agrees with that of $0.150$--$0.389$ from the DEM simulation for a
hopper with $\alpha = 30 \degree$ discharging polydisperse particles
of $\mu=0.3$~\cite{Potapov:96a}. We note that it is reasonable to
compare to the polydisperse system simulated as the monodisperse
system formed hexagonal crystalline structures~\cite{Potapov:96a},
which is not consistant with the physical basis of the current
constitutive model and not realistic for practical hopper flows. This
trend of internal friction increasing from top to bottom is also
supported by DEM simulations of cylindrical flat-bottom
silos~\cite{Langston:95b, Zhu:06c}, where the ratios between the
reported shear and normal stresses varied in a similar
fashion. However, no quantitative comparison can be made due to the
lack of principal stress data and the different geometry used
in~\cite{Langston:95b, Zhu:06c}. With support from the DEM
simulations, our results confirm that the new constitutive model has
advantages in predicting realistic characteristics of stress in the
conical hopper flow.

Predictions of normal stress $\hat{\sigma}_{rr}$ from the present
formulations with and without the NSD effect and from the HGT are
compared in Fig.~\ref{fig:stress_mu03} (d). The HGT uses a constant
$K$ equal to the mean of the $K$ values from the calculation with
NSD. The profiles of $\hat{\sigma}_{rr}$ along the radial distance
$\hat{r}$ are similar, with the HGT result being the largest and
reaching the peak value at the lowest position. The differences in the
normal stress profiles also lead to different predictions of the flow
rate. The values of $\hat{V}$ from the calculations without NSD, with
NSD and using HGT are $1.550$, $1.867$ and $1.992$, respectively,
which shows that larger $\hat{\sigma}_{rr}$ results in a higher flow
rate. To put this flow rate in perspective to commonly measured
discharge rates, we show that the Beverloo equation in the form of $W
= C\rho\phi\sqrt{g(D-k d)^5}$~\cite{Beverloo:61a, Nedderman:82a},
where D is the aperture diameter and $k$ is of order unity, can be
related to $\hat{V}$ as $C = \hat{V}\pi/(4\sqrt{2\sin\alpha})$. The
$C$ measured in experiments takes a value close to
$0.58$~\cite{Nedderman:82a}. However, according to the predictions of
$\hat{V}$, $C$ is at least $1.22$, about twice of the common value for
a $\alpha=30 \degree$ hopper. This over-prediction should mostly be
ascribed to the radial analysis itself, not the constitutive model
though. One major source of error is the assumption of smooth wall in
this analysis. Because of this gross over-prediction, we shall not
assess the capability of the three different analyses regarding
discharge rate prediction. As a summary of the NSD effect, with NSD
the predictions of the normal stress $\hat{\sigma}_{rr}$ and the flow
rate are higher, but the stress ratio $K$ is lower, than without NSD.
\label{}

\begin{figure}
  \centering
      \subfigure[]{
  \includegraphics[width=2.5 in]{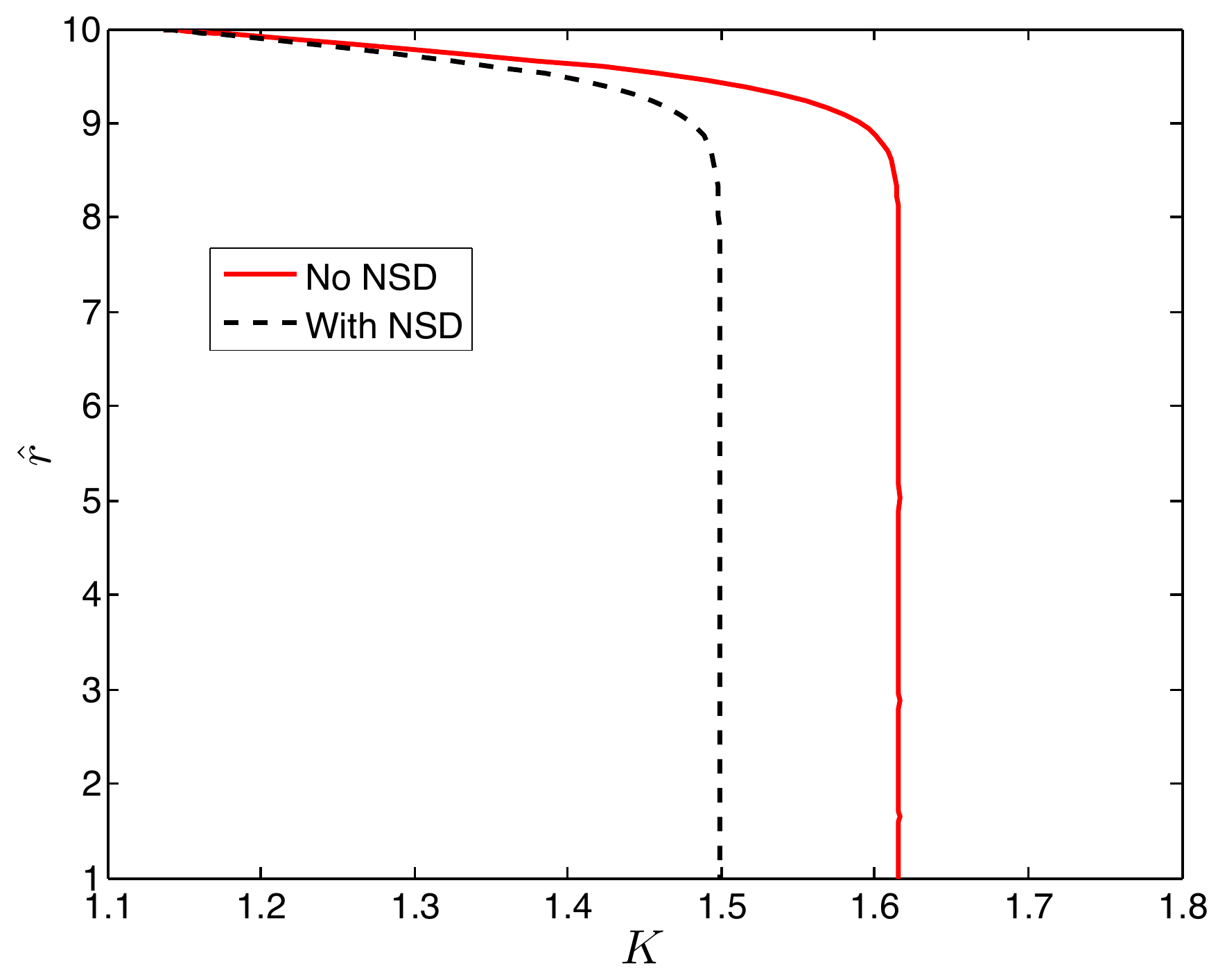}}
      \subfigure[]{
  \includegraphics[width=2.5 in]{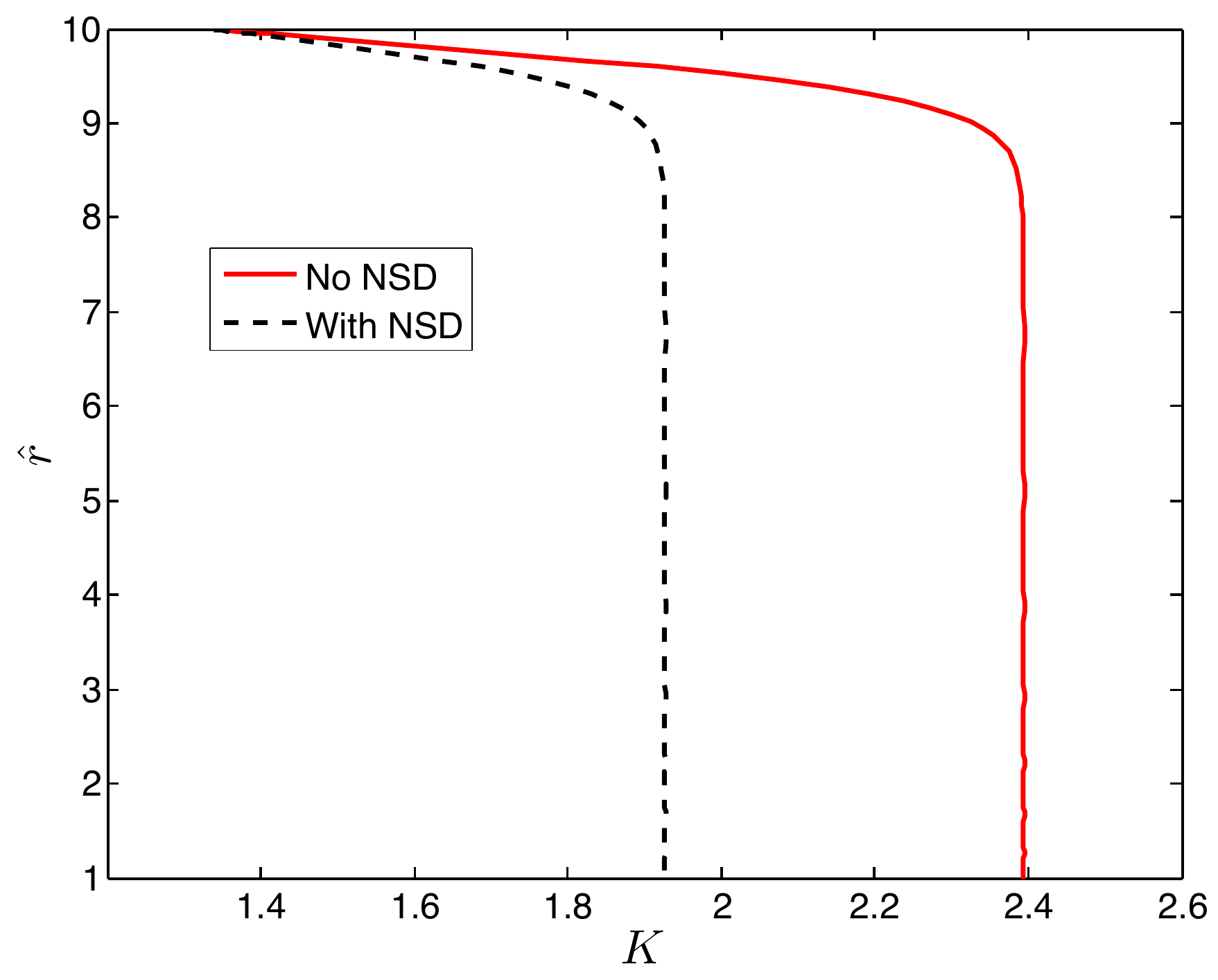}}
      \subfigure[]{
  \includegraphics[width=2.5 in]{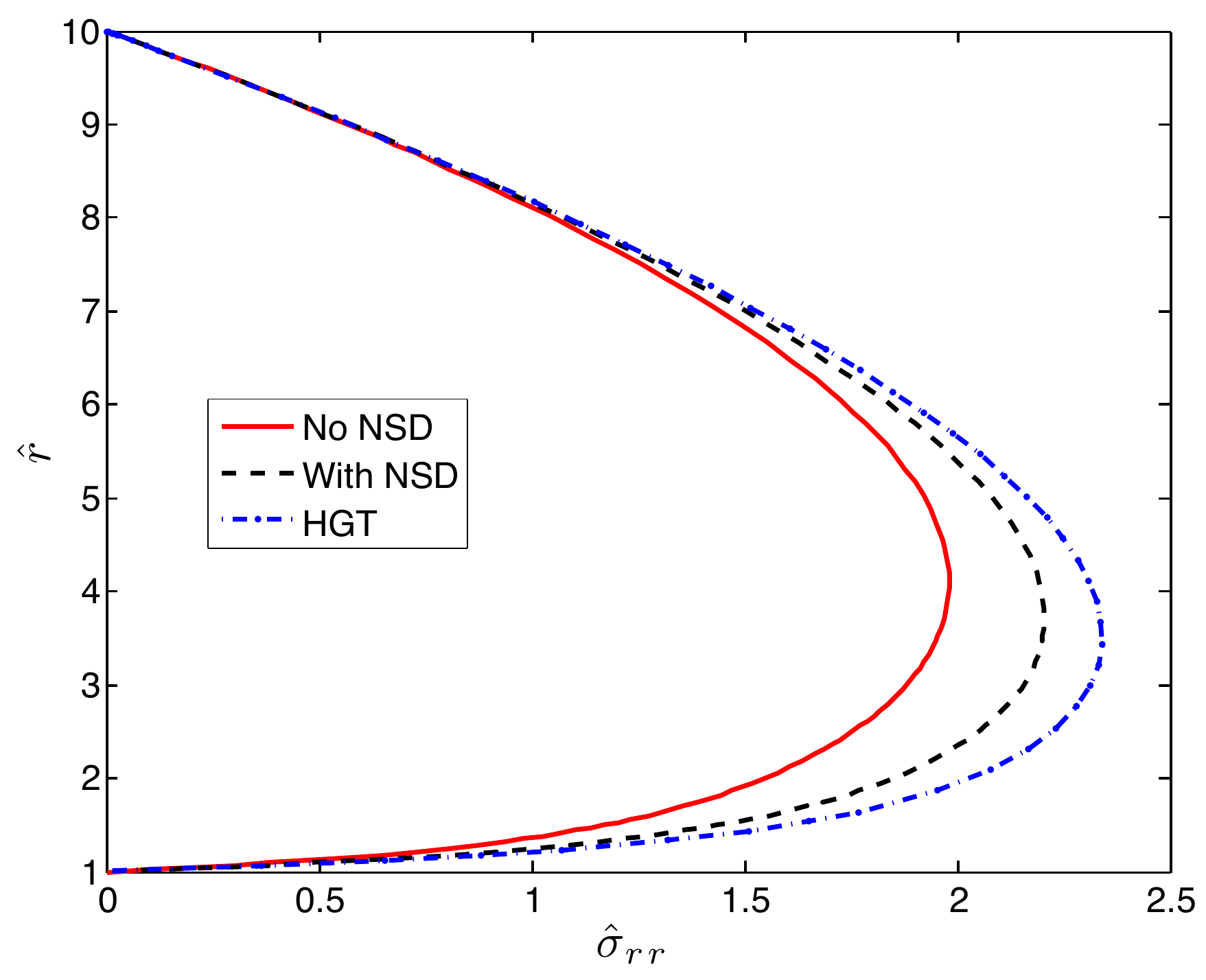}}
     \subfigure[]{
  \includegraphics[width=2.5 in]{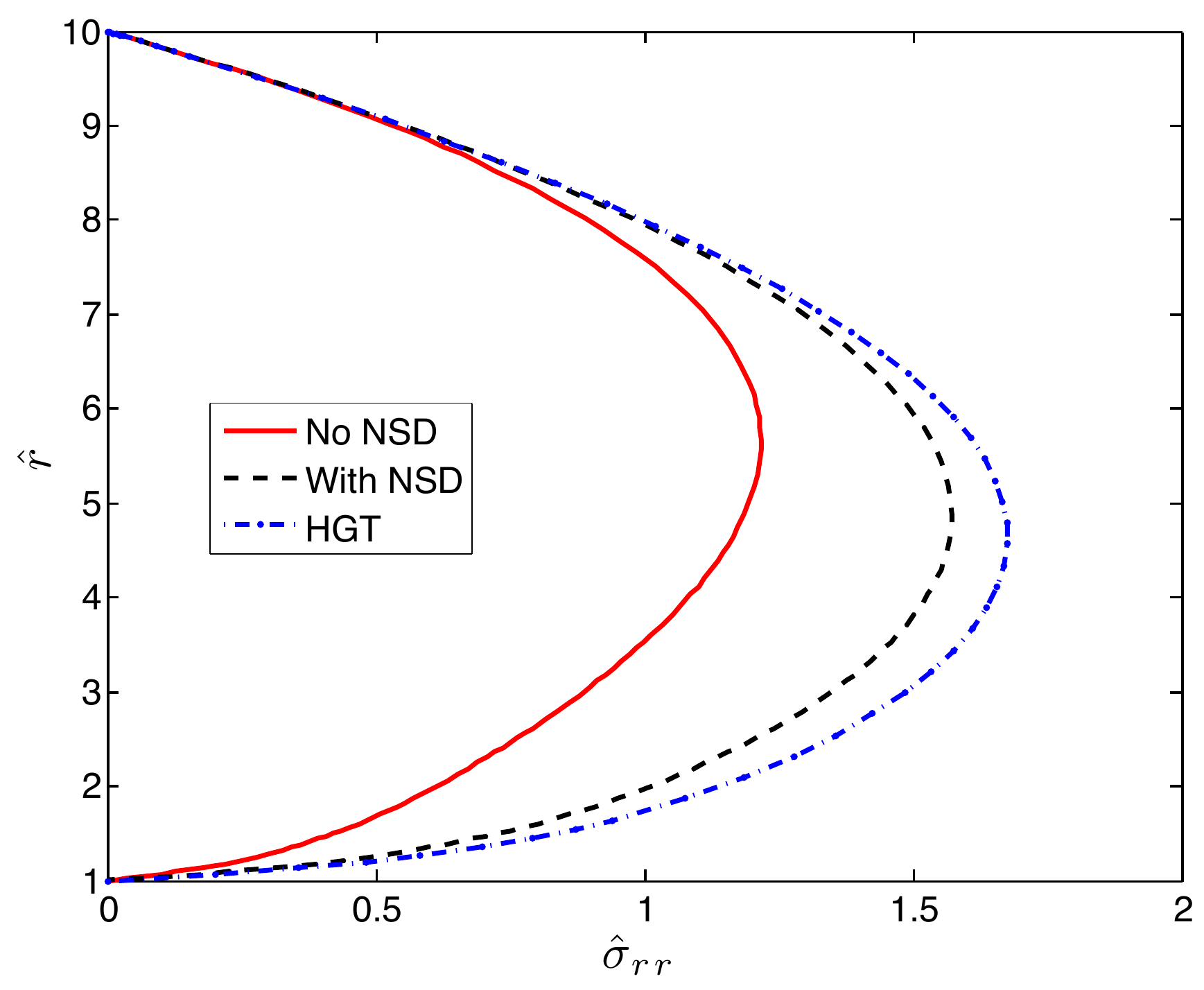}}

\caption{Radial variation of stress ratio in (a) and (b), of normal
  stress in (c) and (d), plotted against the dimensionless distance in
  the vertical axis. The particle-friction coefficient is $\mu=0.1$
  for (a) and (c), and $\mu=1.0$ for (b) and (d). \label{fig:stress_2mus}}
\end{figure}

Another important functionality of the constitutive model used in this
study is to directly probe the particle-friction effect on flow
behavior. We have performed the same hopper-flow analysis for granular
materials with particle friction $\mu=0.1$ and $\mu=1.0$. The results
for $K$ and $\hat{\sigma}_{rr}$ are plotted in
Figs.~\ref{fig:stress_2mus} (a)-(b), and (c)-(d), respectively, of
which (a)-(c) are for $\mu=0.1$ and (b)-(d) for $\mu=1.0$. The fabric
is modeled as independent of $\mu$ and thus not repeated in this
figure. Comparing Figs.~\ref{fig:stress_2mus} (a) and (b) together
with Fig.~\ref{fig:stress_mu03} (a), it can be seen that increasing
$\mu$ results in higher $K$ values. The difference between predictions
with and without NSD is also enlarged. Similar trend is observed for
the internal friction coefficient $f$ (not shown), which is consistent
with engineering observations although no quantitative comparison can
be made at the moment. Studying the $\hat{\sigma}_{rr}$ in
Figs.~\ref{fig:stress_2mus} (c)-(d) and Fig.~\ref{fig:stress_mu03}
(d), it can be concluded that higher $\mu$ leads to lower normal
stress and does not alter the effect of NSD. From the flow-rate
analysis above, it can also be deduced that higher $\mu$ results in
lower flow rates.  Specifically, the $\hat{V}$ values from the
calculations without NSD, with NSD, and using HGT equal $2.161$,
$2.403$ and $2.545$, respectively, when $\mu=0.1$; and $1.365$,
$1.718$ and $1.83$, respectively, when $\mu=1.0$. This effect of
particle friction qualitatively matches practical experience and is
supported by the DEM simulations of a cylindrical
silo~\cite{Hilton:10a}, but again quantitative comparison cannot be
made due to the lack of directly comparable data. The increase in
friction also amplifies the difference between the flow rate
predictions for cases with and without NSD from about $11\%$ to
$26\%$. However, the potential of the constitutive model in
facilitating understanding of the particle-property influence on the
stress distribution and flow rate can clearly be seen from the above
results.

\section{Conclusions \label{sec:conc}}
\label{}
A radial-flow solution of the conical hopper flow has been carried out
utilizing a recently-developed constitutive
model~\cite{Sun:11a}. Results on stress, microstructure and flow rate
have been obtained by simultaneously solving the continuity, momentum
and fabric evolution equations. It has been demonstrated that a
varying stress ratio is predicted, as a result of the fabric
evolution. The particle friction has been shown to lead to higher
stress ratios, but lower normal stress and flow rates as it
increases. These results are supported qualitatively and
quantitatively by DEM simulations and engineering observations. The
effects of considering the NSD, have been explored, showing that
considering NSD results in higher normal stress and flow rate, but
lower stress ratio. The difference can be as large as about $30\%$,
indicating that the NSD in constitutive models warrants more attention
in order to make more accurate predictions.

The significance of the constitutive model has been demonstrated to
lie in connecting particle-level properties to continuum-level
computation through microstructure evolution, obviating the need for
specifying bulk properties and enabling prediction of
particle-property effects on flow. Admittedly, the radial analysis
performed is too simplified to validate against other spatial
measurements, or to be used as a predictive tool for hopper
design. Future work has been planned on applying the constitutive
model to multi-dimensional analysis of compressible flow and
validating against salient hopper flow behavior, such as shear
localization.

\section*{Acknowledgment}
  The authors would like to acknowledge the support from the
  Department of Energy National Energy Technology Laboratory under grants
  DE-FG26-07NT43070 and DE-FE0006932.











\end{document}